\documentstyle[12pt]  {ioplppt}
\begin{document}
\jl{1}
\title{Surface crossover exponent for branched polymers in two
dimensions}[Crossover exponent for branched polymers]
\author{S L A de Queiroz\ftnote{1}{E-mail: sldq@if.uff.br}}

\address{Instituto de F\'\ii sica, UFF, Avenida Litor\^anea s/n,
Campus da Praia Vermelha,\\ 24210--340 Niter\'oi RJ, Brazil}

\begin{abstract}
Transfer-matrix methods on finite-width strips with free
boundary conditions are applied to lattice site animals, which provide a model
for randomly branched polymers in a good solvent. By assigning a distinct
fugacity to sites along the strip edges, critical properties at the special
(adsorption) and ordinary transitions are assessed. The crossover exponent
at the adsorption point is estimated as $\phi = 0.505 \pm 0.015$, consistent
with
recent predictions that $\phi = 1/2$ exactly for all space dimensionalities.
\end{abstract}

\maketitle

\section{Introduction}
The conformational properties of linear polymers near an attractive wall
are well understood by now~\cite{eisen}. The fact that conformal invariance
concepts~\cite{cardy} are applicable to the problem has been extremely helpful,
especially in two dimensions (in which  case the ``wall'' is a line) where
these tools provide a number of exact values for critical exponents. In
contrast, for branched polymers it has been shown that the
underlying field theory is {\it not} conformal~\cite{mdb}.
Exact results on bulk properties of randomly branched polymers have, however,
been obtained through a connection with the theory of Yang-Lee edge
singularities~\cite{ps}. The
corresponding extension towards surface properties has been accomplished only
recently~\cite{jl} yielding, among others, the interesting prediction that the
crossover exponent at the adsorption point has the hyperuniversal
(dimension-independent) value $\phi = 1/2$.
This result applies for an impenetrable wall; penetrable
surfaces have not been considered~\cite{jl}. The case in favour of
hyperuniversality is built from the following elements~\cite{jl}: $(i)$ an
exact calculation in $d = 3$, by means of a
correspondence between branched polymers and an epidemic process plus a
supersymmetric mapping of the
latter onto a semi-infinite one-dimensional Yang-Lee edge problem; $(ii)$
conformal invariance properties of the two-dimensional Yang-Lee problem leading
to information on four-dimensional branched polymers near a surface; $(iii)$
mean-field theory, expected to be valid for $d \geq 8$; and $(iv)$ perturbation
theory in $d = 8 - \epsilon$ dimensions, all four of which  yield $\phi = 1/2$.

In the present work we use finite-size scaling~\cite{fs1} and
phenomenological renormalisation~\cite{fs2} ideas to study surface properties
of site lattice animals (which provide a model for randomly branched polymers
in a good solvent) in two dimensions. To this end, the correlation length for
animals on infinite strips with free boundary conditions (FBC) is numerically
calculated by diagonalisation of the corresponding transfer matrix~\cite{dds}.
By imposing FBC one is enabled to assess surface critical
behaviour, in particular the adsorption transition. Our main goal is to check
on the proposed hyperuniversal value $\phi = 1/2$ at
the adsorption point.
Accordingly, only impenetrable surfaces are considered throughout this paper.
 We start by applying standard one-parameter phenomenological renormalisation
(PR)~\cite{fs2}, reobtaining bulk quantities such as the critical fugacity
$x_c$ and the temperature-like exponent
$y = 1/\nu$ which are very accurately known~\cite{dds,dst}.
This is important as a check of the overall reliability of our procedures.
We then
search for a surface-driven transition, by introducing a distinct fugacity
for occupied sites along the strip boundaries. A two-parameter
PR analysis is carried out, by comparing
correlation lengths on three strips of consecutive widths~\cite{dh,vyj,dqy}.
Two non-trivial fixed points are found, which are respectively related to the
ordinary (bulk-dominated) and adsorption transitions. The corresponding
finite-size estimates of critical parameters and exponents are extrapolated.
Our main final result is $\phi = 0.505 \pm 0.015$, providing support to the
hyperuniversality conjecture~\cite{jl} $\phi = 1/2$ at least in two dimensions.
\section{Model and calculational procedure}

The generating function of the model, defined on a semi-infinite lattice, can
be
written as:
\begin{equation}
{\cal Z} = \sum_{N,N_s} C_{N, N_s} x^N x_s^{N_s} \ ,
\label{eq:1}
\end{equation}
\noindent where $C_{N, N_s}$ is the number of different configurations that
can be built with a total of $N$ sites constrained to form one cluster, of
which $N_s$  are at the surface; $x$ is the fugacity for site occupation and
$x_s = \exp{\epsilon_s/k_{B}T}$ where  $-\epsilon_s$ is the extra energy
assigned to each site at the surface. It is expected on general
grounds~\cite{binder} that
the critical fugacity $x_c$ will be a constant as a function of $x_s$,
from small $x_s$ up to the adsorption threshold given by some $x_s^c > 1$.
Upon approach to a point ($x_c^0$,$x_s^0$) on the critical line in ($x$,$x_s$)
space, the bulk correlation length $\xi$ diverges as $\xi \sim \delta^{-\nu}$
where the scaling field $\delta$ is a  suitable linear combination of
$\delta x \equiv x - x_c^0$ and $\delta x_s \equiv x_s - x_s^0$ . Close to
the adsorption point, a second length $\xi_s$ diverges with a different
exponent
$\nu_s \equiv 1/y_s$ (and a different combination $\delta_s$ of the variables
$\delta x$ and $\delta x_s$). Physically, $\xi_s$ measures the
thickness of the adsorbed polymer layer. Thus the average number of surface
contacts $< N_s >$ scales asymptotically with the average number of sites
$< N >$ as~\cite{binder}
\begin{equation}
< N_s > \ \sim \  < N >^{\phi} \ , \ \ \ \ \ \phi \equiv y_s/y\ .
\label{eq:2}
\end{equation}
Below the adsorption threshold one has $\phi = 0$, while in the adsorbed
phase $\phi = 1$. At the threshold $\phi$ is expected to take on a non-trivial
value.

\medskip

The correlation length $\xi_L(x,x_s)$ along a strip depends on the largest
eigenvalue $\Lambda_L^0\,(x,x_s)$ of the transfer
matrix via~\cite{dds} : $\xi_L (x,x_s) = -1/\ln \Lambda_L^0\,(x,x_s)$.
For large $L$
and close enough to criticality of the corresponding (semi-)infinite
system, the correlation length must scale, in
terms of $\delta$ and $\delta_s$ defined above, as~\cite{fs1}
\begin{equation}
L^{-1}\xi_L(\delta, \delta_s) = F(L^y \delta, L^{y_s} \delta_s) .
\label{eq:3}
\end{equation}
Upon rescaling of $\xi_L$, one expects two non-trivial fixed
points~\cite{vyj,dqy,binder}: $(i)$ the ordinary fixed point,
which governs the critical behaviour of the unbound (bulk) phase, at which
the surface interactions are irrelevant and thus exhibits $y_s < 0$; and $(ii)$
the special fixed point, corresponding to the adsorption transition, with
$y_s > 0$.

We use strips of width $L \leq 10$ sites, both for square and triangular
lattices. Building up the transfer matrix involves the analysis of
connectivity properties of adjacent columns of occupied and empty sites.
For the present case of animals on strips with FBC, this is a straightforward
extension of earlier work on percolation and animals with periodic
boundary conditions (PBC)~\cite{dds,dst} and percolation with FBC~\cite{dq}.
The resulting matrix is rather sparse, owing to restrictions
imposed by connectivity~\cite{dds}: for $L = 10$
on the square lattice, for instance, only $4.2\%$ of the possible combinations
of adjacent column states are allowed.

Extrapolation of finite-width results must be dealt with carefully, especially
as convergence of estimates produced with FBC is usually slower than that of
their counterparts generated with use of PBC~\cite{dq,earlyd,burk}.  In the
present work, extrapolations toward
 $L \to \infty$ have been done using the Bulirsch-Stoer (BST)
algorithm~\cite{bst,ms}. As extensively discussed elsewhere~\cite{ms},
whenever the leading correction-to-scaling exponent $\omega$ is not
known {\it a priori} BST extrapolations rely on keeping it as a free parameter
within an interval guessed to be reasonable.
Central estimates and error bars are evaluated
self-consistently by selecting the range of $\omega$ for which overall
fluctuations are minimised.

\section{Results}

\subsection{One-parameter renormalization}

We first consider no surface binding ($x_s \equiv 1$). We can then
implement standard, one-parameter, PR in the usual way by looking for a
finite-size estimate of $x_c$ given by the fixed point $x^{\ast}_L$ of the
implicit recursion relation:
\begin{equation}
{\xi_L (x^{\ast}_L) \over L} = {\xi_{L-1} (x^{\ast}_L) \over L-1} \ ,
\label{eq:4}
\end{equation}
At the fixed point, an approximation to
the bulk exponent $y = 1/\nu$ is evaluated by~\cite{fs1}:
\begin{equation}
y_L = { \ln \{(d\xi_L/dx)_{x^{\ast}_L}/(d\xi_{L-1}/dx)_{x^{\ast}_L} \} \over
\ln (L/L-1)} - 1\ .
\label{eq:5}
\end{equation}
In order to check on universality of critical amplitudes~\cite{pf}, we also
calculate
the quantity ${\cal A}_L \equiv 2 L /\pi \xi_L (x^{\ast}_L)$.
Note that for a triangular lattice with FBC the strip width $L = N \sqrt{3}/2$,
where $N$ is the number of sites across the strip.
If the underlying field theory were
conformal at the critical point, this would be an estimate of the exponent
describing the decay of critical correlations along the surface, $\eta_s$.

\begin{table}
\caption{
Results from one-parameter PR.  Uncertainties in last quoted digits are shown
in parentheses. Extrapolations obtained by BST algorithm
with correction-to-scaling exponent $\omega$ in ranges shown. Expected values
from Ref. \protect{\cite{dst}}} .
\begin{indented}
\item[]\begin{tabular}{@{}lllllll}
\br
&\centre{3}{Square}&\centre{3}{Triangular}\\ \ns
&\crule{3} & \crule{3}\\
$L$ & $x^{\ast}_L$& $y_L$ & ${\cal A}_L$& $x^{\ast}_L$& $y_L$ & ${\cal A}_L $\\
\mr
3 & 0.298906 &  1.43406  &  1.02271  & 0.247186 & 1.39519 &  0.998368 \\
4 & 0.274596 &  1.45806  &  1.27009  & 0.221892 & 1.42363 &  1.25590\\
5 & 0.263725 &  1.47490  &  1.44619  & 0.210650 & 1.44332 &  1.43950\\
6 & 0.257947 &  1.48691  &  1.57871  & 0.204743 & 1.45791 &  1.57633\\
7 & 0.254533 &  1.49588  &  1.68219  & 0.201287 & 1.46922 &  1.68213\\
8 & 0.252364 &  1.50284  &  1.76528  & 0.199107 & 1.47828 &  1.76637\\
9 & 0.250909 &  1.50841  &  1.83350  & 0.197652 & 1.48570 &  1.83506\\
10& 0.249890 &  1.51296  &  1.89052  & 0.196638 & 1.49190 &  1.89217\\
Expected & 0.246150(10) & 1.5607(4) & --- & 0.192925(10) & 1.5607(4) & --- \\
Extrapolated& 0.2460(2) & 1.55(1) & 2.45(4) & 0.1928(2) & 1.55(1) & 2.4(1)\\
$\omega$    & 1.5(4)  & 1.5(5)& 1.45(20) &1.5(4) &  1.5(5) & 1.5(5)\\
\br
\end{tabular}
\end{indented}
\end{table}

Our results are shown in table 1. Overall agreement with expected values,
where these are available, is rather good. Universality of critical
correlation-length amplitudes is satisfied within error bars. However,
finite-size data show that the amplitude of corrections is much larger than for
the corresponding
cases of PBC (see e.g. table I of reference~\cite{dst}).
Partially as a consequence of this,  our extrapolated estimates for $x_c$ and
$y$ are somewhat less accurate than those obtained with PBC~\cite{dst}.
A second source of imprecision emerges when one considers the broad ranges
allowed for the correction-to-scaling exponent $\omega$ in table 1.
Though some degree of subjectivity is inevitable when dealing with error
estimation within the BST scheme, our results for $\omega$ reflect
the fact that, roughly for $\omega$ between 1 and 2 the fluctuation estimates
for fixed  $\omega$ (based on the spread between next-to-highest order
estimates~\cite{ms}) keep to the same order of magnitude. On the other hand,
outside this interval fluctuations increase, and estimates deteriorate,
quickly. This is to be compared e.g. to
similar extrapolations for percolation with FBC~\cite{dq} where usually one
can pinpoint a much narrower band of values of $\omega$ within which
fluctuations are minimised.

\subsection{Two-parameter renormalisation}

We next allow the surface interaction to vary.
Similarly e.g. to studies of linear polymer adsorption~\cite{vyj,dqy}, an extra
energy $-\epsilon_s$ is introduced for sites on either strip boundary, so that
$x_s = \exp{\epsilon_s/k_{B}T}$. $L$-dependent fixed points
$(x^{\ast},x_s^{\ast})$ are obtained by comparing correlation lengths on
three strips~\cite{dh}:
\begin{equation}
{\xi_L (x^{\ast},x_s^{\ast}) \over L} = {\xi_{L-1} (x^{\ast},x_s^{\ast}) \over
L-1} = {\xi_{L-2} (x^{\ast},x_s^{\ast}) \over L-2} \ .
\label{eq:6}
\end{equation}
In the present case these equations give two fixed points: the ordinary fixed
point,
which describes the behaviour of the unbound animal and the special fixed
point which describes the animal's behaviour at the binding transition.
Linearizing around the fixed points, the exponents $y$ and $y_s$ can be found
from suitable partial derivatives evaluated at the fixed point in
question~\cite{dh,vyj,dqy}. Again we calculate the quantity ${\cal A}_L \equiv
 2 L /\pi \xi_L (x^{\ast},x_s^{\ast})$. Our results for the
ordinary and special fixed points are displayed respectively in tables 2 and 3.

As a general rule, finite-size estimates differ from their limiting ($L \to
\infty$) values by much smaller amounts than was the case in one-parameter PR.
In several instances, though, convergence turns out not to be monotonic.
Further,
within the BST scheme we frequently find the following as the trial value
of $\omega$ is increased from 0.45 to, say, 6: $(i)$
fluctuation estimates at fixed $\omega$ always decrease, and $(ii)$  last-order
approximants ${\cal Q}(\omega )$ vary monotonically, and seem to be converging
towards fixed points(that is, $ d{\cal Q}(\omega )/d\omega \to 0$). This is
consistent with what is found from three-point extrapolations adjusting
$\omega$ for the best straight-line fit of data against
$L^{-\omega}$~\cite{dst}:  as a rule, $\omega$ tends to converge to
unrealistically high values.
Thus, although strictly speaking there are no regions where the BST algorithm
is
stable with
respect to $\omega$~\cite{ms} in such cases, one can produce reasonably
reliable
estimates by looking at trends followed upon increasing $\omega$. For the
entries in tables 2 and 3 to which this picture applies we display the ranges
of
variation of last-order approximants corresponding to
$\omega \geq \omega_{min}$, with $\omega_{min}$ as given in the respective
entry.

 \begin{table}
\caption{
Results from two-parameter PR at the ordinary fixed point.  Uncertainties in
last quoted digits are shown in parentheses. Extrapolations obtained by BST
algorithm
with correction-to-scaling exponent $\omega$ in ranges shown.}
\begin{indented}
\item[]\begin{tabular}{@{}llllll}
\br
\ms
\centre{6}{ (a) Square}\\ 
$L$ & $x^{\ast}$& $x_s^{\ast}$& $y$ & $y_s$ & ${\cal A}_L$\\
\mr
6 &  0.246282 & 0.223963 & 1.57916 & -- 1.02151 & 2.49768\\
7 &  0.246420 & 0.232356 & 1.57746 & -- 1.05163 & 2.48955\\
8 &  0.246352 & 0.225924 & 1.57451 & -- 1.04907 & 2.49477\\
9 &  0.246294 & 0.217824 & 1.57223 & -- 1.04172 & 2.50040\\
10 & 0.246254 & 0.209961 & 1.57057 & -- 1.03480 & 2.50516\\
Expected   & 0.246150(10)$^{\rm a}$ & --- & 1.5607(4)$^{\rm a}$ & -- 1$^{\rm
b}$ & --- \\
Extrapolated & 0.24617(3) & 0.18(1) &  1.566(1) & -- 1.014(6) & 2.519(4) \\
$\omega$ &     $> 3.0$  & $> 3.0$ &  $> 3.0$  & $> 3.0$  & $> 3.0$ \\
\br
\ms
\centre{6}{ (b) Triangular}\\ 
$L$ & $x^{\ast}$& $x_s^{\ast}$& $y$ & $y_s$ & ${\cal A}_L$\\
\mr
6 &  0.194510 & 0.352877 & 1.53424 & -- 0.836625 &  2.381520\\
7 &  0.193555 & 0.289360 & 1.54238 & -- 0.804216 &  2.445904\\
8 &  0.193238 & 0.257802 & 1.54615 & -- 0.841891 &  2.472772\\
9 &  0.193115 & 0.240468 & 1.54833 & -- 0.880553 &  2.485607\\
10 & 0.193056 & 0.229288 & 1.54982 & -- 0.910094 &  2.492939\\
Expected & 0.192925(10)$^{\rm a}$ & --- & 1.5607(4)$^{\rm a}$ & -- 1$^{\rm b}$
& --- \\
Extrapolated & 0.19296(1) & 0.185(15) &  1.555(2) & -- 1.00(1) & 2.512(2) \\
$\omega$ &      $> 3.0$   & ---$^{\rm c}$ &  2.4(9)   & 3.3(5)  &  2.4(6) \\
\br
\end{tabular}
\item $^{\rm a}$Ref. \protect{\cite{dst}}
\item $^{\rm b}$Ref. \protect{\cite{bc}}
\item $^{\rm c}$No optimal $\omega$ found (see text).
\end{indented}
\end{table}

\subsubsection{The ordinary transition.}
For the ordinary transition, the corresponding fixed point can be
found only for $L \geq 6$. The exact result $y_s = -1$ is expected to hold, as
it is based on general properties of the ordinary transition of two-dimensional
systems~\cite{bc}.
For both square and triangular lattices, extrapolations were performed
discarding data for $L=6$.
Though, for the latter, these do not usually imply non-monotonic variation
along the sequence, their
inclusion would increase the scatter of extrapolates by at least one order of
magnitude. For the non-universal $x_s^c$ on the triangular lattice, we have
found neither an optimal range for $\omega$, nor the smooth decrease of error
as $\omega$ increases, described above. Thus we quote for  $x_s^c$ an average
of last-order estimates for $1.0 \leq \omega \leq 4.0$. In general, the final
results for the ordinary fixed point show agreement to within less than
$0.5\%$ with those of Ref.~\cite{dst}; for the exact result
$y_s = - 1$~\cite{bc} fluctuations are higher, but still kept smaller
than $2\%$.
Universality of the amplitude ${\cal A}$ is satisfied within $0.05\%$.

\begin{table}
\caption{
Results from two-parameter PR at the special fixed point.  Uncertainties in
last quoted digits are shown in parentheses. Extrapolations obtained by BST
algorithm
with correction-to-scaling exponent $\omega$ in ranges shown.}
\begin{indented}
\item[]\begin{tabular}{@{}llllll}
\br
\ms
\centre{6}{ (a) Square}\\ 
$L$ & $x^{\ast}$& $x_s^{\ast}$& $y$ & $y_s$ & ${\cal A}_L$\\
\mr
5 &  0.244202 &  2.34075 & 1.54294 & 0.717355 & -- 0.0980272\\
6 &  0.246045 &  2.28235 & 1.55829 & 0.796012 & -- 0.0579771\\
7 &  0.246033 &  2.28278 & 1.56222 & 0.796973 & -- 0.0583358\\
8 &  0.246088 &  2.28049 & 1.56514 & 0.800373 & -- 0.0561538\\
9 &  0.246108 &  2.27951 & 1.56697 & 0.801618 & -- 0.0551068\\
10 & 0.246123 &  2.27877 & 1.56825 & 0.802352 & -- 0.0542317\\
Expected   & 0.246150(10)$^{\rm a}$ & --- & 1.5607(4)$^{\rm a}$ & --- & --- \\
Extrapolated & 0.24615(1) & 2.277(1) &  1.571(2) & 0.804(1) & -- 0.054(2) \\
$\omega$ &     $> 3.0$   & $> 3.0$ & $> 3.9$   & ---$^{\rm b}$  &  $> 3.0$ \\
\br
\ms
\centre{6}{ (b) Triangular}\\ 
$L$ & $x^{\ast}$& $x_s^{\ast}$& $y$ & $y_s$ & ${\cal A}_L$\\
\mr
5 &  0.190915 & 2.85122 & 1.53472 & 0.701013 & -- 0.117316 \\
6 &  0.192174 & 2.78810 & 1.54574 & 0.745358 & -- 0.0887032\\
7 &  0.192573 & 2.76501 & 1.55200 & 0.764248 & -- 0.0761753\\
8 &  0.192735 & 2.75442 & 1.55608 & 0.773683 & -- 0.0695443\\
9 &  0.192813 & 2.74873 & 1.55889 & 0.778886 & -- 0.0655337\\
10 & 0.192856 & 2.74527 & 1.56086 & 0.781982 & -- 0.0628311 \\
Expected   & 0.192925(10)$^{\rm a}$ & --- & 1.5607(4)$^{\rm a}$ & --- & --- \\
Extrapolated & 0.19292(1) & 2.745(5) &  1.5663(5) & 0.7888(5) & -- 0.054(3) \\
$\omega$ &     3.5(6)   & 2.7(4) &  3.6(6)   & 3.5(7)  &  2.0(4) \\
\br
\end{tabular}
\item $^{\rm a}$Ref. \protect{\cite{dst}}
\item $^{\rm b}$No optimal $\omega$ found (see text).
\end{indented}
\end{table}

\subsubsection{The special transition.}
For the special fixed point on the square
lattice we have discarded $L = 5$ and 6 data for  $x^{\ast}$, $x_s^{\ast}$ and
 ${\cal A}_L$ on account of non-uniform convergence; otherwise, all data in
table 3 have been used in extrapolations. For $y_s$ on the square lattice,
fluctuations were approximately constant and small throughout  the range of
$\omega$ explored, so we quote an average
of last-order estimates for $1.0 \leq \omega \leq 4.0$.  While extrapolates
from square-lattice results undoubtedly suffer as a result of the
above-mentioned difficulties, application of the BST algorithm nevertheless
gives a fairly accurate numerical picture.

On the other hand, results for
the triangular lattice fall into smooth, well-behaved sequences from which
we have extracted a set of very precise extrapolates. Our central estimate for
$y$ is higher by $0.4\%$ than that of Ref. \cite{dst}, with non-overlapping
error bars. Recalling that our error bars reflect
uncertainties in the extrapolation procedure itself, and do not take into
account systematic errors in the original sequence of finite-size results, we
do not take this fact as necessarily meaning that our estimates conflict.
Indeed,
other instances are known~\cite{vyj,dqy} in which extrapolates from
two-parameter PR differ slightly e.g. from exact results.

\section{Discussion and conclusions}

It can be seen from tables 2 and 3 that universality of critical
amplitudes~\cite{pf}
is satisfied to within error bars. The qualitative behaviour as one spans
the distinct possibilities is similar to that of the corresponding $\eta$  of
linear polymers.
For PBC ${\cal A} \simeq 0.68$~\cite{dds,dst}; with FBC, at the ordinary
transition ${\cal A} \simeq 2.51$; at the special transition
${\cal A} \simeq - 0.054$. For linear polymers $\eta_{bulk} = 5/24$;
$\eta_s^{Ord} = 5/4$;  $\eta_s^{Sp} = - 1/12$~\cite{vyj,dqy}.
Unfortunately the analogy does not seem to go beyond this level.

The result $x_s^c = 2.277(1)$ for the adsorption threshold on the square
lattice compares well with, and is more precise than, the series estimate
2.25(5)~\cite{dpffs}. It would be interesting to check whether using our
value of $x_s^c$ in the series analysis would improve other results.

Turning now to the crossover exponent $\phi = y_s/y$, the safest course seems
to be separately extrapolating the sequences for $y_s$ and $y$, and then
calculating the ratio of final estimates. From the square-lattice data of
table 2 one would get $\phi = 0.511 \pm 0.002$, while for the triangular
lattice of table 3 one gets $\phi = 0.5035 \pm 0.0005$.
Though, as mentioned earlier, the above error bars do not take into account
systematic errors, it is desirable to have an estimate of these. In order to
do so, we refer to the similar case of adsorption
of two-dimensional linear polymers, where conformal invariance asserts
that $y = 4/3$ and $y_s = 2/3$, thus $\phi = 1/2$ exactly; two-parameter PR
sequences for a locally directed (but globally isotropic) square lattice
extrapolate respectively to $y \simeq 1.339$
and $y_s \simeq 0.679$~\cite{dqy}, which gives $\phi \simeq 0.507$, just under
$2\%$ off the exact value. For fully isotropic lattices, the extrapolation
for $y_s$ overshoots the exact value by $4\%$~\cite{vyj}. In this
latter case the corresponding extrapolations  of $y$ have not been published;
however, finite-lattice data point towards a value somewhat larger than
$4/3$~\cite{vyj}, thus one expects at least a partial compensation of errors
for the ratio $y_s/y$.
 Assuming systematic errors of order $2\%$ in the extrapolated exponents
also for the present case, one gets $\phi = 0.51 \pm 0.01$ (square lattice)
and $ 0.50 \pm 0.01$ (triangular). Systematic deviations are thus estimated to
increase uncertainties by at
least one order of magnitude over those coming from extrapolation procedures.
Our
final result must encompass both the latter central estimates and allow for
the spread between them, plus their own inherent uncertainties. Thus we have
$\phi = 0.505 \pm 0.015$. This is consistent with the hyperuniversality
conjecture $\phi = 1/2$~\cite{jl}, and to be compared with the
series result $\phi = 0.6 \pm 0.1$~\cite{dpffs}.
\medskip

We have studied surface properties of randomly branched polymers in two
dimensions. Our estimates of bulk quantities such as critical fugacity $x_c$
and critical exponent $y$ are in very good agreement with
results obtained with PBC~\cite{dds,dst}.
We have checked that universality of critical amplitudes~\cite{pf} holds in all
instances investigated. The adsorption threshold for the square lattice has
been located with greater accuracy than previously available~\cite{dpffs}.
The crossover exponent $\phi$ at the adsorption point satisfies, within
error bars, the recent hyperuniversality conjecture $\phi = 1/2$~\cite{jl}.

\ack
The author thanks M Henkel for interesting conversations, and
Departamento de F\'\ii sica, PUC/RJ for use of their computational facilities.
This research is supported by CNPq, FINEP and CAPES.

\end{document}